\documentstyle[fleqn,twoside]{article}

\newcommand{\email}[2]{\footnotetext[#1]{e-mail: #2}}

\newcommand{\be}{\begin{equation}}
\newcommand{\ee}{\end{equation}}
\newcommand{\bean}{\begin{eqnarray*}}
\newcommand{\eean}{\end{eqnarray*}}
\newcommand{\bea}{\begin{eqnarray}}
\newcommand{\eea}{\end{eqnarray}}

\def\const{{\rm const}}

\title{The models of delocalized membranes}
\author{M.G. Ivanov${}^1$
 \\
 \em Moscow Institute of Physics and Technology,
  141\,700, Dolgoprudnyi, Moscow Region, Russia}
\date{January 7, 2003}

\begin{document}
\twocolumn[
\maketitle


\begin{abstract}
     The generally adopted approach in theory of relativistic
    strings and membranes, is similar to use of Lagrange 
    coordinates in continious media mechanics.
      One can use an alternative approach, which
    is similar to use of Euler coordinates.
      Under such approach the consideration of 
    thick (delocalized) membranes is natural.
      Membrane kinematics, which coorespond to Euler
    coordinates is constructed.

      Variables, similar to Hamiltonian variables, are
    introduced by means of Legander transformation.
      The case of free membranes appears to be degenerate.

      The examples of exact solutions of Einstein
    equations with delocalized membranes are presented.
\end{abstract}

~

] 
\email 1 {mgi@mi.ras.ru}

\section{Introduction\label{Introduction}}

  This paper present the Euler coordinate approach to membrane theory
 using the results of paper \cite{MGI-DAN,MGI2,MGI-pbr,MGI-Ezh}
 as main examples

  From kinematical point of view membrane is time-like
 submanifold $\bf V$ (world surface) in space-time $M$,
 $\dim{\bf V}=n<D=\dim{\bf M}$.
  Let $\bf V$ is oriented and $\partial{\bf V}=0$.

  One has two natural methods to describe world surface
 $\bf V$
\begin{itemize}
 \item introducing coordinates $\xi=(\xi^0,\dots,\xi^{n-1})$
   at world surface $\bf V$, ${\bf V}=\{X\in{\bf M}| X=x(\xi)\}$;
 \item introducing set of $D-n$ scalar fields
   $\varphi=(\varphi^1,\dots,\varphi^{D-n})$ to
   describe $\bf V$ as level surface
   ${\bf V}=\{X\in{\bf M}|\varphi(X)=\const\}$.
   This approach describes simultaneously all
   membranes corresponding to different level surfaces
   $\varphi=\const$.
\end{itemize}
   Coordinates $\xi$ are related to medium (membrane), so one
 has to consider them as a sort of Lagrange coordinates.
   Coordinates $X$ are related to space-time, so one
 has to consider them as a sort of Euler coordinates.

  Under the first approach one can use the following standard action
 \cite{GSW,BN,VV,Duff}
\be
  S_{standard}[x(\xi)]
  =-T\int\limits_{\bf V} d^n\xi
  \sqrt{\det \gamma_{ij}},
\label{act0}
\ee
 where
$$
  \gamma_{ij}=g_{MN}
   \frac{\partial x^M}{\partial\xi^i}\frac{\partial x^N}{\partial\xi^j}.
$$

  Unfortunately number is Lagrange coordinates $\xi$ is $n$,
 the number of Euler coordinates $X$ is $D$.
  $D>n$, so it is not possible to express Euler coordinates
 in terms of Lagrange ones.

  To make number of Lagrange coondinates equal to $D$ intial
action (\ref{act0}) has to be modified.
  The obvious generalization replaces single membrane $\bf V$
 by set of membranes ${\bf V}(\phi)$, paramerized by
 variables $\phi=(\phi^1,\dots,\phi^{D-n})$
 ({\em simple delocalized membrane})
\be
  S[x(\phi,\xi)]
  =-\int\limits_{\bf F'} d^{D-n}\phi\cdot f(\phi)
    \int\limits_{{\bf V}(\phi)} d^n\xi
  \sqrt{\det \gamma_{ij}}.
\label{act1}
\ee
  Corresponding equations of motion for this action are obviously
 the same as for standard one, but now the analogy of set
 $(\xi,\phi)$ with Largange (related to medium)
 coordinates is transparent.
  Like for standard action, we still have $n$ constrains
 $\frac{\delta S[x(\xi,\phi)]}{\delta x^M}\,\partial_i x^M=0$.
  Let $\frac{D x}{D(\xi,\phi)}\not=0$, then $(\xi,\phi)$
 are new (Lagrange) coordinates at space-time $\bf M$, so
 $\xi=\xi(X)$, $\phi=\varphi(X)$.

  Now one can transform the action from Lagrange coordinates $(\xi,\phi)$
 to Euler coordinates $X=x(\xi,\phi)$.
  New action has to involve integration over
 $X$ instead of $(\xi,\phi)$ and fields $\xi(X)$, $\varphi(X)$
 instead of $x(\xi,\phi)$
\be
  S[\varphi(X)]
 =-\int\limits_{\bf M} d^DX\sqrt{|g|}\,
   f(\varphi)\|d\varphi^1\wedge\dots\wedge d\varphi^{D-n}\|,
\label{Svarphi}
\ee
 where we use norm of $k$-form
 $\|J\|=(J,J)$,
 $(A,B)=\sqrt{\frac1{k!}A_{N_1\dots N_k}B^{N_1\dots N_k}}$.
  Fortunately new action does not involve $n$ fields $\xi(X)$
 and now we have no constraints.

  The action of this form was initially considered by Hosotani et.al.
 \cite{hos1,hos2}.
  Backer and Fairlie \cite{bf1,bf2} suggested an alternative interpretation
 for the same action as action for $(D-n)$-dimensional membrane.
  The same action (\ref{Svarphi}) in the case
 $f(\varphi)=\delta^{D-n}(\varphi)$ was used by Morris and Gee
 \cite{morris1,morris2,morris3} in the context of string
 theory quantisation.
  The action (\ref{Svarphi}) was also used to handle with  Hamilton-Jacobi
 equations for strings and $p$-branes \cite{hos3,bf2}.

\section{Kinematics\label{Kinematics}}
  This section presents more formal view on membrane kinematics
 in Euler (space-time) coordinates.

  Kinematics in Euler coordinates involves two spaces and one
 map from one space, to other one
\begin{itemize}
  \item space-time $\bf M$, $\dim{\bf M}=D$,
   $X^M$ are coordinates on $\bf M$,
   $g_{MN}$ is metric;
  \item auxiliary space $\bf F$, $\dim{\bf F}=D-n<D$,
   $\phi^\alpha$  are coordunates on $\bf F$,
   $\Omega_{\bf F}=f(\phi)\,d\phi^1\wedge\dots\wedge d\phi^{D-n}$
   is volume form;
  \item map $\varphi:\>{\bf M}\to{\bf F}$,
   $\varphi$ is ``potential''.
\end{itemize}
  Using these basic objects we can indroduce the following notions
\begin{itemize}
\item
$$
  \varphi^{-1}(\phi)={\bf V}(\phi),\qquad \dim{\bf V}(\phi)=n<D,
$$
 ${\bf V}(\phi)$ is membrane number $\phi$;
\item
$$
  \varphi^*:\>\Omega_{\bf F}\mapsto
   J=f(\varphi(X))\,d\varphi^1(X)\wedge\dots\wedge d\varphi^{D-n}(X),
$$
 $J$ is ``field intensity'';
\item
 $*J$ is membrane currnet.
\be
  dJ=0\quad\Leftrightarrow\quad \delta(*J)=0
\label{dJ}
\ee
 is ``kinematic conservation law''
 (conservation of membrane current);
\item
\be
  \int\limits_{\bf U}J
\label{flux}
\ee
 is membrane flux through $\bf U$, $\dim{\bf U}=D-n$;
\item
$$
  {\cal G}^{\alpha\beta}
 =g^{MN}\,\partial_M\varphi^\alpha\,\partial_N\varphi^\beta
$$
 is inverse metric generated by $\varphi$ at space
 ${\bf F}'=\varphi({\bf M})$;
\item
 ${\cal G}_{\alpha\beta}$ is matrix inverse to ${\cal G}^{\alpha\beta}$
 (${\cal G}_{\alpha\beta}{\cal G}^{\beta\gamma}=\delta_\alpha^\gamma$);
\item
\be
  \|J\|=f(\varphi)\,\sqrt{\det({\cal G}^{\alpha\beta})}
\label{normJ}
\ee
 is density of membranes in attendant coordinates;
\item
${\rm n}=\frac{J}{\|J\|}$ is unit normal form ($\|{\rm n}\|^2=1$);
\item
 $u=*{\rm n}$ is unit tangent form (``velocity form'', $\|u\|^2=-1$);
\item
$
  \bar P_{MN}
 ={\cal G}_{\alpha\beta}\,\partial_M\varphi^\alpha\,\partial_N\varphi^\beta
 =\frac{{\rm n}_{MK_2\dots K_{D-n}}{\rm n}_{N}{}^{K_2\dots K_{D-n}}}{(D-n-1)!}
$
 is projectior to normal directions of membrane;
\item
\be
  P_{MN}=g_{MN}-\bar P_{MN}
 =\frac{-u_{MK_2\dots K_n}u_{N}{}^{K_2\dots K_n}}{(n-1)!}
\label{P_MN}
\ee
 is projector to tangent directions of membrane
 (projector to membrane world surface).
\end{itemize}

  In some problem it is useful to use attendant (Lagrange)
 coordinates $X=(\xi,\phi)$.
  Metric $g_{MN}$ and inverse metric $g^{MN}$
 are splitted into four blocks:
$$
  g_{MN}=
  \left(
   \begin{array}{cc}
     g_{mn}&g_{m\nu}\\
     g_{\mu n}&g_{\mu\nu}
   \end{array}
  \right),\qquad
  g^{MN}=
  \left(
   \begin{array}{cc}
     g^{mn}&g^{m\nu}\\
     g^{\mu n}&g^{\mu\nu}
   \end{array}
  \right).
$$

  $\gamma_{mn}=g_{mn}$ is metric at membrane world surface ${\bf V}(\phi)$.

  ${\cal G}^{\mu\nu}=g^{\mu\nu}$ is inverse metric at auxiliary
 space $\bf F'$.

  If $\det{\gamma_{mn}}\not=0$, then in attendant coordinates
\be
  g_{\rm attendant}=\frac{\det{\gamma_{mn}}}{\det{{\cal G}^{\mu\nu}}}.
\label{g=gammaG}
\ee
  Equation (\ref{g=gammaG}) allows to derive action (\ref{Svarphi})
 from (\ref{act1}).

  Identity is (\ref{g=gammaG}) is trivial consequence of the
 following statement:

  {\em Let $G$ is $N\times N$-matrix, $\det G\not=0$.
$$
  G=
  \left(
   \begin{array}{cc}
     A&B\\
     C&D
   \end{array}
  \right),\qquad
  G^{-1}=
  \left(
   \begin{array}{cc}
     K&L\\
     M&N
   \end{array}
  \right),
$$
 where $A$, $B$, $C$, $D$ are matrices
 $n\times n$, $n\times(N-n)$, $(N-n)\times n$ and $(N-n)\times(N-n)$
 respectively. $K$, $L$, $M$, $N$ are matrices of the same
 dimensions as $A$, $B$, $C$, $D$.
  If $\det A\not=0$, then 
$$
 \det G=\frac{\det A}{\det N}.
$$
}

\section{Dynamics\label{Dynamics}}

  In this section we continue the analogy above between
 membrane theory in Euler coordinates and classical mechanics
 for dynamics.

  Let us consider the action
\be
  S[\varphi(X)]=\int\limits_{\bf M}d^DX\sqrt{|g|}\,
    L(\varphi,J).
\ee
  One can introduce ``momentum'' $K$ by following formula
\be
  K=\frac{\partial L}{\partial J}.
\ee
  Variation of action gives
\be
  \frac1{\sqrt{|g|}}\frac{\delta S}{\delta\varphi^\alpha}=
  \frac{\partial L}{\partial\varphi^\alpha}-(J_{(\alpha)},\delta K),
\ee
 where
\be
  J_{(\alpha)}=(-1)^{D+n+\alpha}
   f(\varphi)\,d\varphi^1\wedge\dots\widehat{\wedge d\varphi^\alpha}
    \dots\wedge d\varphi^{D-n}.
\ee
  The term under the hat has to be skipped.

  By means of Legendre transformation one can
 define ``Hamiltonian''
\be
  H_{\rm mem.}(\varphi,K)=(K,J(\varphi,K))-L(\varphi,J(\varphi,K));
\ee
\bea
  \frac{\partial H_{\rm mem.}}{\partial K}&=&J,\\
  \frac{\partial H_{\rm mem.}}{\partial\varphi^\alpha}
   &=&-(J_{(\alpha)},\delta K).
\label{H-eqn}
\eea
  System (\ref{H-eqn}) of $D-n$ equations contains not only $\varphi^\alpha$
 and $J$, but also $\partial_M\varphi^\alpha$.

  For some application it would be useful to modify system (\ref{H-eqn})
 in following way.
$$
  \frac{\partial H_{\rm mem.}}{\partial\varphi^\alpha}\partial_M\varphi^\alpha
   =-(J_{(\alpha)},\delta K)\partial_M\varphi^\alpha=-(J,\delta K)_M.
$$
  Here we introduce new notation
 $(J,\delta K)_M=\frac1{(D-n-1)!}J_{K_1\dots K_{D-n-1}M}
  (\delta K)^{K_1\dots K_{D-n-1}}$.
  If $\|J\|\not=0$, then matrix $\partial_M\varphi^\alpha$ has rank $D-n$,
 i.e. new system is equivalent to (\ref{H-eqn}).
  Taking into account
\bean
  \nabla_MH_{\rm mem.}
 &=&
  \frac{\partial H_{\rm mem.}}{\partial\varphi^\alpha}\partial_M\varphi^\alpha
 +\left(\frac{\partial H_{\rm mem.}}{\partial J},\nabla_MJ\right)\\
 &=&
  \frac{\partial H_{\rm mem.}}{\partial\varphi^\alpha}\partial_M\varphi^\alpha
 +(K,\nabla_MJ)
\eean
 we write
\be
  \nabla_MH_{\rm mem.}=(K,\nabla_MJ)-(J,\delta K)_M.
\label{H-eqn2}
\ee
  System (\ref{H-eqn2}) is equivalent to (\ref{H-eqn}).

  For simple delocalized membrane $L(\varphi,J)=-\|J\|$, $K=-{\rm n}$,
 where ${\rm n}=\frac{J}{\|J\|}$,
 so $J$ is not expressable in terms of $K$ (information on $\|J\|$ does
 not included to ${\rm n}$),
 and ``Hamiltonian''
 does not exist.
  I.e. simple delocalized membrane is degenerate case.

  All sections below deals with membranes of this sort, which
 are descibed by the following equations of motion 
\be
  ({\rm n},\delta{\rm n})_M=0.
\ee

\section{Localization\label{Localization}}

   To reproduce the case of unit tension single free membrane
 with world surface $a(x)=0$ one has to set
\be
   J=\delta(a)\,da^1\wedge\dots\wedge da^{D-n}.
\label{Jsing}
\ee
   There are two obvious possibilities to reproduce this form of $J$.
  $\delta$-function can originate from form of action
  (from function $f(\phi)$) or from form of field $\varphi$
\begin{enumerate}
  \item set $f(\phi)=\delta(\phi)$, $\varphi=a$ \cite{morris1,morris2}, or
  \item set $f(\phi)=1$, $\varphi^\alpha=\theta(a^\alpha)$ \cite{MGI-DAN}
\end{enumerate}

  Field (\ref{Jsing}) is singular, but in the case of
 simple delocalized membrane (\ref{Svarphi}) one could
 consider equations of motion $({\rm n},\delta{\rm n})_M=0$
 for field of this form, because to calculate
 $({\rm n},\delta{\rm n})_M$
 in some point of membrane one need only values of ${\rm n}$ along
 the same membrane.

\section{Energy\label{Energy}}

   For simple delocalized membrane (\ref{Svarphi}) energy-momentum tensor
 has the form
$$
 T_{MN}=-\rho P_{MN},
$$
 where $\rho=\|J\|$ is density of membrane
 matter (\ref{normJ}), and $P_{MN}$ is projector to membrane world surface
 (\ref{P_MN}).

  Equation of motion for free membrane is equivalent to conservation
 of energy-momentum tensor $\nabla_MT^{MN}=0$.

  Hamiltonian density (it is not ``Hamiltonian'' from section \ref{Dynamics})
 in Minkowski space-time with signature $(-,+,+,\dots,+)$ is
\be
  {\cal H}=\sqrt{\det(\partial_m\varphi^\alpha\,\partial_m\varphi^\beta)+
     \partial_m\varphi^\alpha\,\partial_m\varphi^\beta\,p_\alpha p_\beta},
\ee
 whete index $m=1,\dots,D-1$ numerate space coordinates,
 $p_\alpha$ are canonical momenta.

\section{Generalized Majumdar-Papapetru solution\label{Generalized_M-P}}

  In this section we build explicitly simplest delocalized
 $p$-brane solution, the generalized Majumdar-Papapetru solution.
  It describes static charged dust cloud with charge density equal
 (in ``natural'' units) to mass density.
  In classical Majumdar-Papapetru solution, instead if dust, one
 has set of charged extremal black holes with charge equal to mass.

  Delocalized $p$-brane solutions for arbitrary $p$ in arbitrary
 dimensions were presented in papers \cite{MGI-DAN,MGI2}.
  In paper \cite{MGI-pbr} intersecting delocalized $p$-brane
 solutions were built (intersecting $p$-brane solutions 
 with no delocalised sources are reviewed in papers
 \cite{AIR,IM,IM-rev})

  Let us consider Einstein-Maxwell action with dust source
$$
  S=\int d^4X\sqrt{-g}
    \left(
      \frac{R}2-\frac{\|F\|^2}{2}
     -\|J\|+\frac1{\sqrt2}\,(*J,A)
    \right),
$$
 where $A$ is electromagnetic potential (1-form),
 $F=dA$ is electromagnetic field,
 $J=d\varphi^1\wedge d\varphi^2\wedge d\varphi^3$
 is 0-brane (dust) field.
  $\frac1{\sqrt2}*J$ is 4-dimensional current density.

  By variation over fields $g_{MN}$, $A_M$ and
 $\varphi^\alpha$ one finds the following equations
 of motion
\be
  R_M^N-\frac{R}2\delta_M^N=F_{MK}F^{NK}-\frac{\|F\|^2}2\delta_M^N
  -\|J\|P_M^N,
\ee
$$
  \left(d\frac{*J}{\|J\|}+
 \frac{F}{\sqrt2}\right)\wedge d\varphi^\beta\wedge d\varphi^\gamma=0,
 \quad
  \delta F=-\frac{*J}{\sqrt2}.
$$
  The fields defined by the following equations solve the
 equations of motion
$$
 ds^2=-H^{-2}dX^0dX^0
 +H^2\delta_{\alpha\beta}dX^\alpha dX^\beta,
$$
$$
  A=\frac{\sqrt2}H\,dX^0,
 \quad
  J=-2\triangle H\,dX^1\wedge dX^2\wedge dX^3.
$$
  Here $H$ is smooth positive function
 of space coordinates $X^\alpha$,
 $\alpha,\beta=1,2,3$,
 $\triangle=\partial_1^2+\partial_2^2+\partial_3^2$.
  Calculating $\|J\|$ one has to choose a branch
 of square root, which gives dust density
 $\|J\|=-2H^{-3}\triangle H$.

\section{Black hole with radial strings\label{Black_hole}}

  Black holes with radial string were recently
 considered in papers \cite{FFP,MGI-Ezh}
 (more detailed bibliography is available in these papers),
 where some very general  results were presented.
  In this section we present the simplest black hole
 solution with arbitrary continious distribution of strings.
  More general case of charged black hole and
 muldidimensional case were considered in paper
 \cite{MGI-Ezh}.

  Let us consider the following action
\be
  S=\int_{\bf M}d^4X\sqrt{-g}
    \left(\frac{R}2-\|J\|\right),
\label{act4DJ}
\ee
 where 2-form $J=d\varphi^1\wedge d\varphi^2$ is string field.
  Norm $\|J\|$ represents density of string matter.

  By variation of fields $g_{MN}$ and $\varphi^\alpha$
 ($\alpha=1,2$) one can find equations of motion
\be
  \label{eomM}
  R^M_N-\frac12\delta^M_NR=-\|J\|P^M_N,\qquad
  d\frac{*J}{\|J\|}\wedge d\varphi^\alpha=0.
\ee
  Let space-time ${\bf M}$ is warped product with warp factor $r^2$
 of 2D black hole ${\bf B}$:
$$ds_{\bf B}^2=-\left(1-\frac{2M}{r}\right)dt^2+\frac{dr^2}{1-\frac{2M}{r}},$$
 and an arbitrary 2D Riemannian manifold
 ${\bf S}$ with metric $ds^2_{\bf S}=e^{2f(u,v)}(du^2+dv^2)$,
 i.e. $ds^2=ds_{\bf B}^2+r^2\,ds^2_{\bf S}$.
  Every metric of this form is a solution of equations of motion (\ref{eomM})
\bea
  ds^2&=&-\left(1-\frac{2M}{r}\right)dt^2
       +\frac{dr^2}{1-\frac{2M}{r}}
\nonumber
\\
      &&{}+r^2e^{2f(u,v)}\,(du^2+dv^2),
  \label{cont}
\\
\nonumber
  J&=&({\cal K}-1)\,\Omega_{\bf S},
\eea
 where $\Omega_{\bf S}=e^{2f(u,v)}\,du\wedge dv$ is 2-form of volume 
 at ${\bf S}$, and
\be
  {\cal K}=-e^{-2f(u,v)}\triangle f(u,v)
\ee
 is a half of Riemannian curvature of ${\bf S}$
 ($\triangle=\partial_u^2+\partial_v^2$).
  In definition of norm $\|J\|$ one have to choose branch of square root,
 which corresponds to $\|J\|=\frac{{\cal K}-1}{r^2}$, so
 the case ${\cal K}<1$ corresponds to negative string tension.

  This solution represents black hole with horizon at $r=2M$.
  Every surface $u=\const$, $v=\const$ is string world surface.
  The density of string distribution is $\|J\|$.
  Number (total tension) of string intersecting some
 2-dimensional surface ${\bf U}$ is determined by equation
 (\ref{flux}).

  For compact ${\bf S}$ the total string tension is (\ref{flux})
\be
  \mu_{\bf S}=\int\limits_{\bf S}J=2\pi\chi({\bf S})-A({\bf S}),
\ee
 where $\chi({\bf S})$ is Euler characteristic of ${\bf S}$,
 and $A({\bf S})>0$ is area of ${\bf S}$.
  If one require string tension to be positive, then
 $\mu_{\bf S}>0$, and $\chi({\bf S})>0$ is necessary
 condition on topology of ${\bf S}$.


\section{Conclusion}

  The paper presents a wide class of models of delocalized membranes
 using Euler coordinate approach.
  In Euler coordinates one has only bulk equations and no worldsurface ones.
  Euler coordinate approach involves no constraints, it
 allows to build Hamiltonian formalism in a straightforward
 manner, which could be useful for quantization.

  Simple delocalized membrane is a degenerate case, which has
 the most direct relation to commonly used Lagrange coordinate
 approach.

  Membrane delocalization is useful as method of removing
 of singular sources from field equations.
  To recover singular sources one could consider appropriate limit.
  Nevertheless this procedure is rather nontrivial.
  E.g. classical Majumdar-Papapetru solution has horizons,
 meanwhile generilized Majumdar-Papapetru solution has no gorizons.
  Another difficulty is related with definition of delocalized
 membrane equations of motion in points with zero or infinite
 membrane density.
  It is interesting that the difficulties mentioned have not
 only mathematical, but also physical interpretation.
  The existing of physical viewpoint enhance our intuition
 in handling with delocalized membrane models.

\section*{Acknowledgement}
 {The author is grateful to I.V. Volovich,
 M.O. Katanaev and V.N. Melnikov.
  The work was partially supported by grant
 RFFI 99-01-00866.
  The author is grateful to INTAS for travel support.}


\small
\begin{thebibliography}{99}

\bibitem{GSW}
  Green M., Schwarz J.H., Witten E.,
  Superstring theory, vol.1,2,
  Cambridge University Press, 1987

\bibitem{BN}
 {Barbashov B.M., Nesterenko V.V.,}
 Model' relyativistkoj struny v fizike adronov,
 Moscow.: Energoatomizdat, 1987. 176 p. (in Russian)

\bibitem{VV} {V.S. Vladimirov, I.V. Volovich},
 Doklady Akademii Nauk. 1985. vol. 289. pp. 1043--1047

\bibitem{Duff}{Duff M.J., Lu J.X.},
  Black and super p-branes in diverse dimensions,
  Nucl.Phys. B416, 1994. pp. 301--334; hep-th/9306052

\bibitem{hos1} Y. Hosotani,
  Local scalar fields equivalent to Nambu-Goto strings,
  Phys.Rev.Letters 47 (1981) 399

\bibitem{hos2}
 R.I. Nepomechie, M.A. Rubin, Y. Hosotani, 
 A new formulation of the string action,
 Phys.Lett.B105 (1981) 457

\bibitem{hos3}Y. Hosotani and R. Nakayama,
  The Hamilton-Jacobi Equations for Strings and p-Branes
  Mod.Phys.Lett. A14 (1999) 1983;
  hep-th/9903193
\bibitem{morris1} T.R. Morris,
 From First to Second Quantized string Theory,
 Phys.Lett. 202B (1988) 222
\bibitem{morris2} D.L. Gee and T.R. Morris,
  From First to Second Quantized string Theory. 2.
 The Dilaton and other Fields,
  Nucl. Phys. B331 (1990) 675
\bibitem{morris3} D.L. Gee and T.R. Morris,
  From First to Second Quantized string Theory. 3.
 Gauge Fixing and Quantization,
 Nucl.Phys. B331 (1990) 694

\bibitem{bf1} L.M. Baker and D.B. Fairlie,
  Companion Equations for Branes,
  Journal of Mathematical Physics 41 (2000) 4284--4292;
 hep-th/9908157
\bibitem{bf2} L.M. Baker and D.B. Fairlie,
 Hamilton-Jacobi equations
  and Brane associated Lagrangians,
 hep-th/0003048

\bibitem{AIR}
 {Aref'eva I.Ya., Ivanov M.G., Rytchkov O.A.},
 Wess J., Akulov V.P.(Eds.),
 Supersymmetry and Quantum Field theory,
 Proceedings of the D.~Volkov Memorial Seminar
 Held in Kharkov, Ukraine, 5--7 January 1997,
 Springer, pp. 25--41; hep-th/9702077

\bibitem{IM}
 {Ivashchuk V.D., Melnikov V.N.},
 P-brane black holes for general intersections,
 Grav.Cosmol. 5, 1999. pp. 313-318; gr-qc/0002085
\bibitem{IM-rev}
 {V.D. Ivashchuk, V.N. Melnikov},
 Exact solutions in multidimensional gravity
  with antisymmetric forms,
 Class. Quantum Grav. 18 (2001) R1-R66; hep-th/0110274

\bibitem{FFP} V.P. Frolov, D.V. Fursaev and D.N. Page,
 Thorny spheres and black holes with strings,
 hep-th/0112129

\bibitem{MGI-DAN} {M.G. Ivanov},
 The model of delocalized membranes,
 Doklady Akademii Nauk. 2001. vol. 378. pp. 26--28
\bibitem{MGI2}{M.G. Ivanov,}
 Delocalized membrane model,
 hep-th/0105067; Grav.Cosmol. Vol.8(2002), No.3(31), pp.166--170
\bibitem{MGI-pbr}{M.G. Ivanov},
 Intersecting delocalized p-branes,
 hep-th/0107206
\bibitem{MGI-Ezh}{M.G. Ivanov},
 Black holes with complex multi-string configurations,
 hep-th/0111035; Grav.Cosmol. Vol.8(2002), No.3(31), pp.171--174

\end{thebibliography}
\end{document}